\documentclass[12pt]{amsart}
\usepackage{amsmath, amssymb,latexsym}
\input epsf

\theoremstyle{plain}
\newtheorem{theorem}{Theorem}[section]
\newtheorem{lemma}[theorem]{Lemma}

\theoremstyle{definition}
\newtheorem{definition}[theorem]{Definition}
\newtheorem{condition}[theorem]{Condition}
\newtheorem{problem}[theorem]{Open problem}
\newtheorem{example}[theorem]{Example}

\theoremstyle{remark}
\newtheorem{remark}[theorem]{Remark}

\newcommand{\<}{\langle}
\renewcommand{\>}{\rangle}
\renewcommand{\(}{\left (}
\renewcommand{\)}{\right )}
\newcommand{\A}{{\mathcal A}}
\newcommand{\B}{{\mathcal B}}
\newcommand{\C}{{\mathcal C}}
\newcommand{\F}{{\mathcal F}}

\newcommand{\G}{{\mathcal G}}
\newcommand{\g}{{\mathfrak g}}
\newcommand{\h}{{\mathfrak h}}
\renewcommand{\k}{{\kappa}}
\newcommand{\N}{{\mathbb N}}
\newcommand{\x}{{\vec x}}
\newcommand{\D}{{\mathfrak D}}
\newcommand{\op}[1]{\operatorname{#1}}
\newcommand{\DES}{\op{DES}}
\newcommand{\XOR}{{\sc xor}}
\newcommand{\xor}{\oplus}
\newcommand{\mappedto}{{\stackrel{f}{\mapsto}}}
\newcommand{\Item}[1]{\item{\ {#1}}}
\long\def\forget#1\forgotten{}

\begin{document}

\title[Generators with guaranteed diversity]{Guaranteeing the diversity of number generators}

\author{Adi Shamir}
\address{Department of Applied Mathematics,
Weizmann Institute of Science, Rehovot 76100, Israel}
\email{shamir@wisdom.weizmann.ac.il}
\author{Boaz Tsaban}
\address{Department of Mathematics,
Bar-Ilan University,
Ramat-Gan 52900, Israel}
\email{tsaban@macs.biu.ac.il}
\urladdr{http://www.cs.biu.ac.il/\~{}tsaban}

\begin{abstract}
A major problem  in using iterative number generators of the
form $x_i=f(x_{i-1})$ is that they can enter unexpectedly short
cycles. This is hard to analyze when the generator is designed,
hard to detect in real time when the generator is used, and can
have devastating cryptanalytic implications. In this paper we
define a measure of security, called \emph{sequence diversity},
which generalizes the notion of cycle-length for non-iterative
generators. We then introduce the class of counter assisted
generators, and show how to turn any iterative generator (even a
bad one designed or seeded by an adversary) into a counter
assisted generator with a provably high diversity, without
reducing the quality of generators which are already
cryptographically strong.
\end{abstract}


\maketitle

\section{Introduction}

In this paper we consider the problem of generating long
cryptographically secure sequences by iterative number generators which
start at some seed value $x_0=s$, and extend it by computing
$x_{i}=f(x_{i-1})$ where $f$ is some function. The $i$th output of
the generator is a (typically shorter) value $y_i=g(x_i)$ derived
from the internal state by some output function $g$ (Figure 1). If $f$ is a
secret keyed function, then $g$ may be the identity.

\begin{figure}[!h]
\begin{displaymath}
\epsfysize=3 truecm {\epsfbox{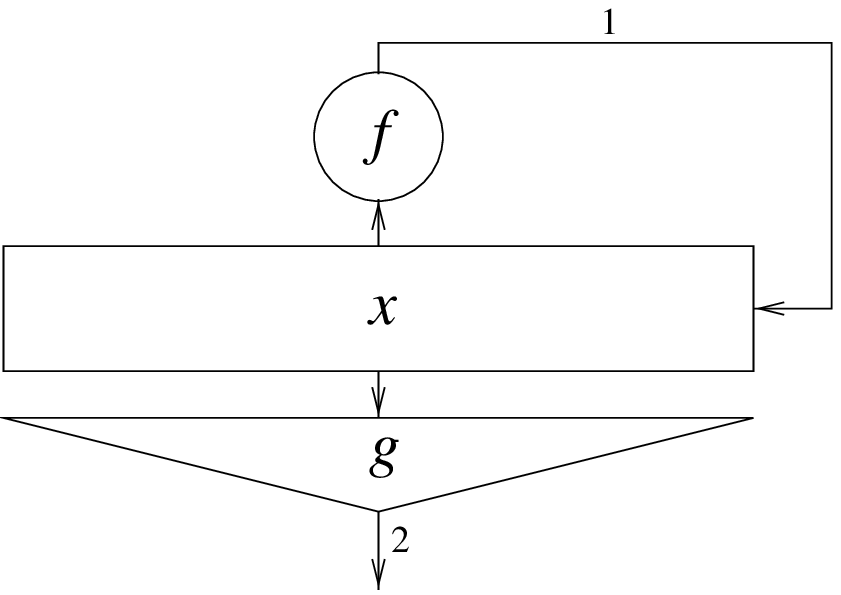}}
\end{displaymath}
\caption{}
\end{figure}

A major application of number generators is to encrypt
cleartexts by \XOR{}ing them with the generated outputs.
In this case, the seed $s$ is a secret key which is shared by the communicating parties,
but is unknown to the eavesdropping adversary.

Since the state space is finite, the sequence of internal states $x_i$ will
eventually become periodic with some period $p$, i.e.,
$x_{i}=x_{i+p}$ for all $i$ larger than some $i_0$. Any cycling of
the state sequence causes a cycling of the output sequence with
period \emph{at most} $p$. A particularly worrisome problem is the
possibility that $i_0$ and $p$ may be unexpectedly small, and
therefore the cycling
point $i_0+p$ is actually achieved.
This can happen even in very complex generators. An interesting
example is Knuth's ``Super-random'' number generator (Algorithm
{\bf K}) \cite[\S 3.1]{KNUTH}, which converges rapidly to a fixed
point (that is, $i_0$ is very small, and $p=1$).

If the cycling point $i_0+p$ is achieved, then the \XOR{} of the
$i$th and $i+p$th ciphertexts is equal to the \XOR{} of the $i$th and
$i+p$th cleartexts, for all $i\geq i_0$. If the cleartexts have a
sufficiently high redundancy, the cryptanalyst can detect the
cycling by noticing the non-uniform statistics of such \XOR{}'s, and
then recover the actual cleartexts from their known pairwise
\XOR{}'s. Even if the cleartexts have no redundancy, knowledge of
some cleartexts will make it possible to find other cleartexts
encrypted with the same repeated values.

\subsection{Partial solutions}\label{partial}

\subsubsection{Online monitoring}
A possible solution to this problem is to monitor each execution
in real time. If a particular seed leads to early cycling, the
cryptographic operation is stopped and the seed is replaced.
However, this can be very disruptive if the exchange of new seeds
is time consuming or difficult to arrange. Note further that real
time detection of cycling behavior using hash tables requires a
very large memory, whereas other methods such as Floyd's two pointer cycle detection
algorithm (see, e.g., \cite[p.~7]{KNUTH}) are not guaranteed to
detect cycles as soon as they are entered.

\subsubsection{Experimental testings}\label{testing}
The designer of the generator can test its behavior by applying
$f$ a limited number of times to a limited number of random seeds
(see \cite{ANDERSON}).
However, such testing cannot be exhaustive, and thus even if no
cycling is ever detected in these tests, the next seed or the next
step can lead to a cycling.

\subsubsection{Pseudorandom functions}
\emph{Pseudorandom functions} $f:X\to X$
are functions which are chosen from the space of
all possible functions $g:X\to X$ with a relatively low-entropy
distribution, but which are difficult to tell apart from
\emph{truly random} functions (which are
selected from the space of all possible functions $g:X\to X$
with uniform distribution). For any adversary with unlimited
computational power and access to
a polynomial (in $\log|X|$) number of values of a pseudorandom function
$f$, the probability that the adversary can tell that these values
came from $f$ rather than from a truly random $g$ should be negligible.
\emph{Pseudorandom permutations} and \emph{pseudorandom sequences} are defined similarly to be
low-entropy but difficult to distinguish from truly random
permutations and sequences, respectively. For more precise definitions
see, \cite{Y}, \cite{GGM}, \cite{LUBY}, \cite[\S 2.2]{NAORE}, and references therein.

It is easy to see (and well known) that sequences generated by iterative number generators
with pseudorandom functions $f$ are pseudorandom. Thus,
the probability that such a generator enters a small cycle is negligible.
However, all known constructions of pseudorandom functions are
slow and are based on unproved conjectures (see \cite[\S17.9]{S}).
In fact, all practical functions used in cryptography
are ad-hoc constructions which are not proved
to be pseudorandom, and nothing is known about the actual
structure of the cycles they generate.\footnote{
A notable exception appears in \cite{KJE} and \cite{COHED}, where the
cycle structure of nonlinear feedback shift registers is studied.
However, the obtained results cover only degenerate cases.
Moreover, in \cite{KJE} it is proved that the studied generators
\emph{must have} short cycles.
}
This is particularly worrisome for the user, since there is no guarantee that
the generators that he uses do not contain a trapdoor leading to short cycles.%
\footnote{Knuth's example could be viewed as such a trapdoor
generator. }

\subsubsection{Mathematically structured generators}\label{mathgen}
The need to avoid short cycles is the major motivation behind the
development of several families of generators based on mathematical
structures. These families include:
Linear congruential generators,
linear feedback shift registers (LFSR's),
clock-controlled LFSR's,
additive generators,
feedback with carry shift registers,
$1/p$ generators (see \cite[\S\S 16--17]{S} and references therein),
and TSR's \cite{TV}.
Under certain conditions, these families can be proved to have large cycles.

The drawback of this approach is that their mathematical structure
can be often used to cryptanalyze them (see \cite[loc.\ cit.]{S} for
references to cryptanalysis of various implementations of the mentioned generators).

\subsubsection{Re-keying}
Chambers \cite{CHAMBERS} suggested a technique to
reduce the risk of short cycles by restarting the generator's
internal state every fixed number of iterations, with a new key
seed taken from a ``re-keying'' generator which has a provably
large cycle (e.g., one of the generators mentioned in Section \ref{mathgen}).

Given an iterative generator, let $p_k$, $k=1,2,\dots$,
be the probability that the cycling point
of the generator occurs after at least $k$ iterations.
Assume that we use the generator to get an output sequence of size $m$.
The probability that we do not reach the cycling point in the usual iterative mode
is $p_m$. Now, if we re-key the generator every $k$ iterations, then the
probability that we do not reach the cycling point even once is $p_k^{m/k}$.
As nothing is known on the cycle structure of the generator, there is no guarantee that
$p_k^{m/k}$ is greater than $p_m$. It may thus be the case that the re-keying mode
is worse than the standard iterative mode.


Moreover, if the re-keying generator is cryptographically weak, then
it could be cryptanalyzed from the outputs which come immediately after the re-keying phases.

One should note further that, as Schneier points out in \cite[\S 17.11]{S},
algorithms that have a long key setup routine are not suitable for this mode.

\subsubsection{Similarity transformations and counter-mode}\label{similarity}
Another possible solution is to take some simple permutation $u$
which is guaranteed to have long cycles (e.g., $u(x)=x+1 \pmod
n$, or any of the examples from Section \ref{mathgen}), and then
to use $fuf^{-1}$ (instead of $f$) as the iteration function.
This similarity transformation has the same cycle structure as $u$.

Such a construction is, though, rather degenerate. Let $\<f,g\>$
stand for a generator whose iteration function is $f$, and whose
output function is $g$.
Consider a generator of the form $\<fuf^{-1},g\>$.
Define $\tilde g=g\circ f$. Then for all seeds $s$, setting
$\tilde s=f^{-1}(s)$ implies that the $i$th output is
$g((fuf^{-1})^i(s))=g(fu^if^{-1}(s))=(g\circ f)(u^i(\tilde s))=\tilde g(u^i(\tilde s))$,
that is, the generator is equivalent to the generator $\<u,\tilde g\>$.
This means that the modified generator is equivalent to another
generator with a cryptographically weak iteration function.

For $u(x)=x+1 \pmod n$ we conclude that for some $\tilde g$, the $i$th
output of the generator equals $\tilde g(\tilde s+i)$. Generators
of the form $y_i = g(s+i)$ are called \emph{counter-mode}
generators, and are a standard mode of operation \cite[\S 9.9]{S}.
However, such generators have the following unpleasant property:
The difference of any two input values $s+i$ and $s+j$ to $g$ is
simply $i-j$. If $i$ is close to $j$, then $i-j$ has a small
Hamming weight. This fact could be used in differential or
correlation cryptanalysis of $g$. This is also the case for other
choices of $u$, e.g., if $u$ is an LFSR, then $u^i(s)$ and
$u^j(s)$ are equal in all except for $i-j$ bits.

\section{The diversity of sequence generators}

In this section we propose a new notion of security for sequence
generators, which generalizes the cryptographically desirable
concept of long cycles.

We first define the notion of diversity for a single infinite
sequence.

\begin{definition}
The \emph{diversity of a sequence} $\x=(x_0,x_1,x_2,\ldots)$ is the
function $\D_\x(k)$ for $k=1,2,3,\ldots$ defined as the minimum
number of distinct values occurring in any contiguous subsequence
$x_i,x_{i+1},\ldots,x_{i+k-1}$ of length $k$ in $\x$.
\end{definition}

All of the sequences considered in this paper have a finite sample
space of $|X|=n$ possible values. For any sequence $\x$ in $X$,
\begin{displaymath}
1 \leq \D_\x(k) \leq \D_\x(k+1) \leq \D_\x(k)+1 \leq n.
\end{displaymath}
In other words, the diversity grows monotonically and at most
linearly with $k$, and cannot exceed $n$.

We now generalize the concept from sequences to generators.
We first define the types of generators considered in this paper:

\begin{definition}
An \emph{iterative generator} is a structure $\G = \<X, Y, f:X\to X,
g:X\to Y \>$, where for all $x\in X$, $f(x)$ and $g(x)$ can be
computed in polynomial time from $x$. $X$ is \emph{the state space},
and $Y$ is \emph{the output space}. We may write $\G=\<f,g\>$ for
short, or $\G: x_i = f(x_{i-1})$ if the output function is
not relevant. For a generator $\G: x_i = f(x_{i-1})$ and seed $s\in X$,
we denote the \emph{state sequence} $(x_0=s, x_1, \dots)$ of
the generated internal states by $\G(s)$.
\end{definition}

We wish to bound from below the diversity of the sequences of
internal states generated from possible seeds.

\begin{definition}\label{diver}
The \emph{diversity of an iterative generator} $\G: x_i=f(x_{i-1})$
is the function
\begin{displaymath}
\D_\G(k) = \min\{\D_{\G(s)}(k) : s\in X\}
\end{displaymath}
defined for $k=1,2,3,\ldots$.
The \emph{total diversity} of $\G$ is the limit
$\lim_{k\rightarrow\infty}\D_\G(k)$.%
\footnote{
Anderson, et.\ al., \cite{ANDERSON} suggested a statistically-oriented
notion of diversity for random number generators,
based on experimental testings of the generator. These testings give
estimations for the \emph{average case} behavior, whereas
our notion bounds the \emph{worst case} behavior of the generator.
Moreover, the combinatorial nature of our notion will make it possible
to use mathematical theory in order to
apply it to cases where experimental testings are not suitable
(e.g., when the state space is huge). See also Section \ref{testing}.
}
\end{definition}


Iterative generators on finite spaces have simple diversity functions.

\begin{lemma} Assume that $\G: x_i=f(x_{i-1})$ is an iterative generator.
\begin{enumerate}
\Item{Let $\x$ be a sequence (of internal states) created by $\G$.
Then $\D_\x(k)=\min\{k,p\}$ where $p$ is the length of the cycle
that $\x$ enters into.}
\Item{$\D_\G(k) = \min\{k,p\}$ where $p$ is the length
of the shortest cycle in $f$.}
\end{enumerate}
\end{lemma}

\begin{proof}
$\x$ has distinct values before it enters the cycle and while it
completes the first traversal of the cycle. This implies (1), and
(2) follows from (1).
\end{proof}
\forget $X$ has distinct values before it enters the cycle and
while it completes the first traversal of the cycle, since
otherwise it would enter an earlier or shorter cycle.
Consequently, any contiguous subsequence of length $k<p$ in $X$
contains exactly $k$ distinct values. When $k \geq p$,
subsequences starting on the cycle cannot have more than $p$
distinct values. If $p'$ is the length of the shortest cycle in
$f$, then any contiguous subsequence of length $k<p'$ on any $X$
computed by $f$ contains $k$ distinct values, and subsequences
starting on the shortest cycle cannot have more than $p'$ distinct
values. Consequently, $\D_G(k)=min(k,p')$.
\end{proof}
\forgotten

The diversity of an iterative generator is thus directly related
to the size of its smallest cycle.
It is intended to capture one aspect of the worst case
behavior of a generator, in the sense that generators with
provably high diversity cannot repeat a small number of internal
states a large number of times
as a result of an unlucky or adversarial choice of seed.


The diversity measure can be applied to noniterative generators,
in which the computation of $x_i$ may depend on its index $i$ as well.

\begin{definition}
A \emph{counter-dependent} generator is a structure
$\G = \<X, Y, F:X\times \N\to X,g:X\to Y \>$,
where for all $x\in X$ and $i\in\N$, $F(x,i)$ and $g(x)$ can be
computed in polynomial time from $x$. $X$ is \emph{the state space},
and $Y$ is \emph{the output space}.
In this type of generators, the next state is calculated by $x_i = F(x_{i-1},i)$.
Here too, we denote the \emph{state sequence} $(x_0=s, x_1, \dots)$ of
generated internal states by $\G(s)$.
\end{definition}

Note that iterative as well as counter-mode generators are particular cases of
counter-dependent generators.
A straightforward generalization of Definition  \ref{diver} for counter-dependent generators is:
\begin{definition} \
\begin{enumerate}
\item{The diversity of a counter-dependent generator $\G : x_i = F(x_{i-1},i)$
is the function $\D_\G(k) = \min\{\D_{\G(s)}(k) : s\in X\}$
defined for $k=1,2,3,\ldots$.
The \emph{total diversity} $\D_\G^\mathrm{total}$
of $\G$ is the limit
$\lim_{k\rightarrow\infty}\D_\G(k)$.}
\item{A counter-dependent generator $\G : x_i = F(x_{i-1},i)$ is \emph{$\g(k)$-diverse}
if $D_\G(k)\geq\g(k)$ for all $k=1,2,\ldots$.}
\end{enumerate}
\end{definition}

The diversity of a general counter-dependent generator can grow and
freeze in an irregular way when $k$ increases, since these
generators are not forced into a cycle when they accidentally
repeat the same $x_i$ value. The diversity function is thus a
natural generalization of the notion of cycle size.

\section{Modifying generators}

In this section we consider several ways in which we can modify a
given iterative generator in order to increase its diversity.  The
main intuitive conditions we impose on this process are:

\begin{condition}\label{same}
{\rm
We do not want to design the new generator from scratch. We
usually prefer to use known and well studied primitives such as
DES, RC5 or nonlinear feedback shift registers, for which highly
optimized code can be easily obtained or reused from other parts
of the application. We thus want the modified design to
use the same cryptographic ingredients as the original design.
}
\end{condition}

\begin{condition}
{\rm
The computational complexity of the modified next-state function must not be significantly
greater that that of the original one.
}
\end{condition}

\begin{condition}
{\rm
The modification technique should be uniformly applicable to all
iterative generators, treating them as black boxes. We do not want
the modification to be based on the mathematical or statistical
properties of the given iteration function $f$. In particular, we
can not assume that we know the structure of its cycles.
}
\end{condition}

\begin{condition}
{\rm
We are more interested in increasing the diversity of the interval values $x_i$ than
in increasing the diversity of the output values $y_i=g(x_i)$:
If the given generator uses an output function $g$ with a small
range (e.g., a single bit) applying diversity measures to the
output values is meaningless.
}
\end{condition}

The modification should be a win/win situation: If the given
generator has a low diversity, the problem should be rectified,
but if the given generator is already strong, we do not want the
modification to weaken it.
The problem is that we do not have a
general quantitative definition of the ``goodness'' of generators,
except when they are ``perfect''. We thus concentrate in this
paper on the following formal interpretation.

\begin{condition}\label{win-win} \
{\rm
\begin{enumerate}
\Item{For any given iteration function, the modified generator should be
$\g(k)$-diverse for some $\g(k)$ which is exponential in $\log n$.}
\Item{If the iteration function $f$ is pseudorandom, then
the state sequences generated from random seeds by
the modified generator should be pseudorandom.}
\end{enumerate}
}
\end{condition}

As in counter-mode (see Section \ref{similarity}), our black box
modification technique is based on turning the iterative generator
into a counter-dependent generator, allowing $x_i$ to depend on $i$ in
addition to $x_{i-1}$.
To sharpen our intuition, let us consider some \emph{bad} constructions.
(In the following examples and throughout the paper,
the state space $X$ is identified with the set $\{0,1,\ldots,n-1\}$,
and addition in the state space is carried modulo $n$.)

\begin{example}\label{index}
$x_{i}=i$. This function has maximal diversity, but poor
cryptographic quality.
\end{example}

\begin{example}
$x_i=f(i)$. This is the standard counter-mode.
Perfect generators remain perfect, but for a constant $f$ the
diversity is $1$.
\end{example}

\begin{example}
$x_i=f(i)+i$. This is a simple combination of the previous two
examples. Perfect generators remain perfect,
but for $f(x)=-x$, all the generated $x_i$ are 0, and thus the
diversity is $1$.
\end{example}

\forget
\begin{example}
$x_i=f(x_{i-1})+1$. For $f(x)=x-1$, the original generator has
maximal diversity, and the modified generator has a diversity of
$1$.
\end{example}
\forgotten

\begin{example}
$x_i=f(x_{i-1}+i)$. This is an attempt to force the next state to
depend both on the previous state and on the index. Perfect
generators remain perfect, but the generated sequence has
diversity 1 when $f$ is a constant function.
\end{example}

\begin{example}
$x_i=f(x_{i-1}+i)+i$. This is the ``kitchen sink'' approach,
trying to combine all the ingredients in all possible ways.
However, when the function $f$ is $f(x)=-x$, the sequence
generated from any initial seed $x_0=s$ is $s,-s,s,-s,s,-s,\ldots$
which contains at most two values.
\end{example}

Considering these counterexamples, the reader may suspect that all
black box modifications are bad (for some $f$). In the next
section we show that this is not the case.


\section{A provably good modification technique}\label{provgood}

Given an iterative generator $\<f,g\>$, we apply the following
black-box modification.
\begin{definition}
A \emph{counter-assisted generator} $\<f,g\>$ is a generator in which
$x_0=s$, and for all $i \geq 1$ $x_i=f(x_{i-1})+i \pmod n$,
where $n$ is the size of the state space,
and the $i$th output is $g(x_i)$ (see Figure 2).
\end{definition}

\begin{figure}[!h]
\begin{displaymath}
\epsfysize=4 truecm {\epsfbox{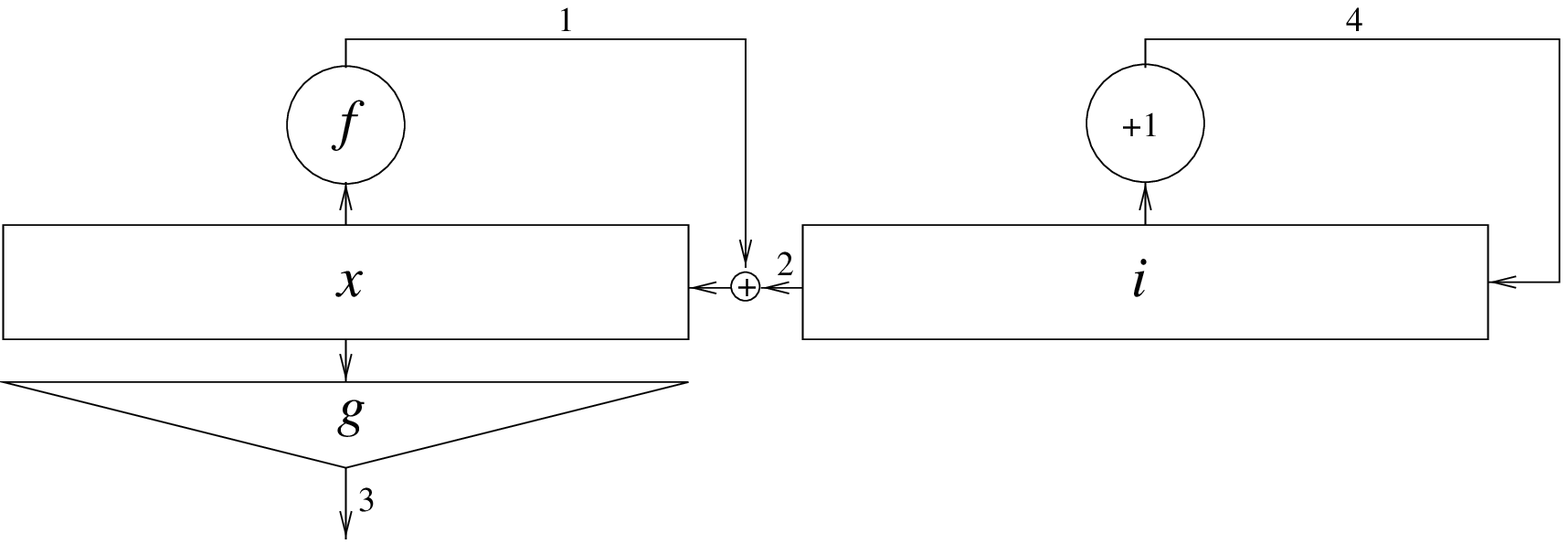}}
\end{displaymath}
\caption{}
\end{figure}

Since it is easy to maintain or obtain a counter for the number of values
produced so far (in many applications, one can use either the loop counter or the running block-number as a counter
for the counter-assisted mode),
and no change is made in the function $f$ or $g$,
the modification technique is completely trivial and can be
applied to any iterative generator without increasing its
complexity.

Formally, for all generators $\<X,Y,f,g\>$, the counter assisted
modified generator is in fact the iterative generator $\<X\times \{0,\dots,n-1\}, Y, F, G\>$, where
\begin{eqnarray}
F(x,i) & = & (f(x)+i \pmod n, i+1 \pmod n)\nonumber\\
G(x,i) & = & g(x)\nonumber\\
\end{eqnarray}
\noindent However, note that:
\begin{enumerate}
\item{The only secret part is located in the $x$ coordinate,}
\item{incrementing $i$ has no cryptographical significance, and}
\item{the output calculation $G(x,i)$ is independent of the $i$-coordinate.}
\end{enumerate}
Thus applying diversity measures on the whole state space $X\times \{0,\dots,n-1\}$---that is,
measuring the diversity of the sequences of pairs $(x_i,i)$, $i=1,2,\dots$---is misleading (and, in fact,
not informative).
This is why the diversity measure is focused on the actual state sequences $\G(s)=(x_0=s, x_1, \dots)$
rather than on the sequence of pairs $(x_i,i)$.

\begin{lemma}\label{isolated-equality}
Let $\x=(x_0,x_1,x_2,\ldots)$ be a state sequence of a counter assisted generator. Then
for all $i \neq j \pmod n$, if
$x_i=x_j$ then $x_{i+1} \neq x_{j+1}$ and $x_{i-1} \neq x_{j-1}$.
\end{lemma}

\begin{proof}
We argue modulo $n$. By definition, $x_{i+1}=f(x_i)+(i+1)$ and
$x_{j+1}=f(x_j)+(j+1)$. If $x_i=x_j$ but $i \neq j$, then
necessarily $x_{i+1} \neq x_{j+1}$.
Now, for the very same reason, $x_{i-1}=x_{j-1}$ would imply
$x_i\neq x_j$, which is not the case.
\end{proof}

In other words, the sequence $\x$ has the interesting property
that equality at any pair of locations implies inequality at the
pair of their immediate successors and the pair of their immediate
predecessors. We call this \emph{the isolated equality property}.
This is the intuitive reason why counter assisted generators
cannot enter short cycles: If they accidentally generate the same
value at several locations, all the subsequent computations are
guaranteed to diverge rather than converge.


\begin{theorem}\label{counter-assisted is good}
\
\begin{enumerate}
\Item{The black box modification technique modifying
$\G: x_i=f(x_{i-1})$ to $\G': x_i=f(x_{i-1})+i \pmod n$
is $\max\{\g(k),\h(k)\}$-diverse, where
\begin{displaymath}
\g(k) =
\begin{cases}
\sqrt{k-1} & k\le n\\ \sqrt{n}   & n < k
\end{cases},
\quad\textrm{and}\quad
\h(k) =
\begin{cases}
k/|\op{Im}(f)| & k\le n\\ n/|\op{Im}(f)|   & n < k
\end{cases}.
\end{displaymath}
}
\Item{If the iteration function $f$ is pseudorandom, then
the state sequences generated from random seeds
by the modified generator are pseudorandom.}
\end{enumerate}
\end{theorem}

\begin{proof}
(1) We first show that $\g(k)\le \D_{\G'}(k)$ for all
$k=1,2,\dots$. Consider any sequence of $k$ consecutive values
$x_i,x_{i+1},\ldots,x_{i+k-1}$ ($k\le n+1$), and assume that it
contains exactly $\nu$ distinct values. There are $\nu^2$ possible
ordered pairs of these values $(a,b)$, and by
Lemma \ref{isolated-equality} each one of them can occur at most
once in a consecutive pair of locations $(x_j,x_{j+1})$ along the
sequence. Since there are $k-1$ such locations, $\nu^2 \geq k-1$,
which yields the desired lower bound on $\nu$.

Next, we need to show that $\h(k)\le \D_{\G'}(k)$ for all
$k=1,2,\dots$. In a sequence of $k$ consecutive values
$x_i,x_{i+1},\ldots,x_{i+k-1}$ ($k\le n+1$), each $x_j$ is of the
form $c_j+j$, where $c_j\in\op{Im}(f)$. Since we add $k$ distinct values
to at most $|\op{Im}(f)|$ values, we get at least $k/|\op{Im}(f)|$
distinct values.

(2) We now sketch the proof of the pseudorandomness part.
Consider the following sequence of oracles, which accept
a number $k$ (which is polynomial in $\log n$)
and output a sequence $x_1,\dots,x_k\in X$. (By \emph{random} we mean
statistically independent and uniformly distributed.)
\begin{itemize}
\item[{\bf Oracle 1:}]{Returns a random sequence $x_i\in X$ ($i=1,2,\dots,k$).}
\item[{\bf Oracle 2:}]{Chooses a random seed $x_0=s$, and defines an $f:X\to X$
\emph{on the fly}, as follows:
\begin{enumerate}
\item A flag ${\sf Birthday}$ is initially set to $0$.
\item For each $i=1,2,\dots,k$:
\begin{itemize}
\item If $f(x_{i-1})$ is undefined, then choose a random $y\in X$ and define $f(x_{i-1})=y$.
\item Otherwise, set ${\sf Birthday} = 1$.
\end{itemize}
\item Set $x_i = f(x_{i-1})+i$.
\end{enumerate}
The remaining values of $f$ are chosen randomly.
}
\item[{\bf Oracle 3:}]{
Chooses a particular function $f$ with uniform probability
from the set of all functions from $X$ to $X$, chooses a random seed $x_0=s$,
and returns the sequence $x_i$ with $x_i = f(x_{i-1})+i$,
$i=1,2,\dots,k$.
}
\item[{\bf Oracle 4:}]{Same as Oracle 3, but with $f$
\emph{pseudorandom} instead of truly random.}
\end{itemize}
We say that two oracles are \emph{distinguishable} if there exists a
(not necessarily polynomial time) algorithm (called \emph{distinguisher})
which, for some constant $c>0$, given a sequence of length polynomial in
$\log n$, can tell with probability greater than $1/\log(n)^c$ which oracle
has generated this sequence. Otherwise, the oracles are \emph{indistinguishable}.
It is clear that Oracles 2,3 are indistinguishable. That Oracles
3,4 are indistinguishable follows from the fact
that any distinguisher of these oracles can be
used to construct a distinguisher of pseudorandom
functions from random ones.

It remains to show that Oracles 1,2 are
indistinguishable. The only possible constraint on the output of
Oracle 2 happens when $f$ is applied twice to the same argument,
that is, $\sf Birthday$ is set to 1. It is well-known that for
$k<<n$, the probability that no birthday occurs is close to
$\frac{k^2}{2n}$ \cite{BIRTHDAY}, which is negligible if $k$ is polynomial in
$\log n$.
\end{proof}

\begin{remark}
The upper bound $\frac{k^2}{2n}$ on the distinguishing probability
is tight: In probability close to $\frac{k^2}{2n}$, a birthday
$x_i=x_j$ occurs and the distinguisher can check that
$x_{i+1}-(i+1)=x_{j+1}-(j+1)$. Provided this, the probability that
the output came from Oracle 1 is $1/n$.
\end{remark}

\section{Asymptotic tightness of the provable diversity}

The square root lower bound
on the diversity may seem to be an artifact of the proof
technique.
We first consider the purely combinatorial version of the problem:
What is the longest sequence one can construct from $\nu$ distinct
symbols which has the isolated equality property?

\begin{lemma}
For any positive integer $\nu$, there exists a sequence of length
$\nu^2+1$ consisting of $\nu$ symbols and having the isolated
equality property.
\end{lemma}
\begin{proof}
Let $C$ be a complete directed graph with $\nu$ vertices and
$\nu^2$ directed edges (including self loops). As the graph is
connected and the indegree and outdegree of each vertex in $C$ is
the same ($=\nu$), the graph is Eulerian. Let $v_0e_0v_1e_1\dots
v_{\nu^2-1}e_{\nu^2-1}v_0$ be an Eulerian tour, which includes
each directed edge exactly once.
Assume that for some distinct $i$ and $j$, $v_i=v_j$.
If $v_{i+1}=v_{j+1}$, then necessarily $e_i=e_j$, which is
disallowed in Eulerian tours. Similarly, $v_{i-1}=v_{j-1}$ would
imply $e_{i-1}=e_{j-1}$. Consequently, the sequence has the
isolated equality property.
\end{proof}
This combinatorial result does not rule out the possibility that
sequences created by counter assisted generators must satisfy
additional constrains, and as a result
the lower bound in Theorem \ref{counter-assisted is good} can be improved
significantly.
We will show that this is not the case:
We prove the asymptotic
tightness of our lower bound by constructing for each $n$
a specific counter-assisted generator, such that the total diversities of
these counter-assisted generators are $O(\sqrt{n})$.

\begin{theorem}\label{tightness}
There exist functions $f_n$, $n=1,2,\ldots$ 
such that the total diversities $\D_{\G_n}^\mathrm{total}$ of
the counter assisted generators
$\G_n: \ x_i=f_n(x_{i-1})+i \pmod n$
are $O(\sqrt{n})$.
\end{theorem}

\begin{proof}
Fix a natural number $n$. We will write for short $f$ and $\G$ instead of
$f_n$ and $\G_n$, respectively.

The state sequence of $\G$ will be based on two sequences:
$a_0, a_1, \ldots, a_{\alpha-1}$ and $b_0,b_1,\ldots,b_{\beta-1}$
(the values of $\alpha$ and $\beta$ will be determined later).
The sequences are ``meshed'' as follows:

\begin{enumerate}
\item{Locations with even indices contain only the
       $a_i$ values, and
       locations with odd indices contain only the $b_j$ values.
}
\item{The $a_i$ values occur in block order: The first $\beta$
       occurrences are $a_0$, the next $\beta$ occurrences are $a_1$,
       and so on.
}
\item{The $b_j$ values occur in cyclic order: The first
       $\beta$ occurrences are $b_0, \ldots,$$b_{\beta-1}$ in this order, the
       next $\beta$ occurrences are again $b_0, \ldots, b_{\beta-1}$ in this order,
       and so on.
}
\end{enumerate}

Putting these blocks in consecutive rows, we get a matrix
$C=(c_{ij})$ of size $\alpha\times 2\beta$,
where $c_{i,2j}=a_i$ and $c_{i,2j+1}=b_j$:
\begin{displaymath}
C =
\begin{pmatrix}
a_0 & b_0 & a_0 & b_1 & \cdots & a_0 & b_{\beta-1} \\
a_1 & b_0 & a_1 & b_1 & \cdots & a_1 & b_{\beta-1} \\
\vdots & \vdots & \vdots & \vdots &    & \vdots & \vdots \\
a_{\alpha-1} & b_0 & a_{\alpha-1} & b_1 & \cdots & a_{\alpha-1} & b_{\beta-1} \\
\end{pmatrix}
\end{displaymath}

We define a function $f$ for which the counter assisted generator $\G: \ x_i=f(x_{i-1})+i$,
seeded by $x_0=a_0$, has state sequence equal to our meshed sequence.

We begin with a few simple restrictions on our parameters.
For cyclicity the counter must return to $0$ after $2\alpha\beta$ steps, that is,
$2\alpha\beta=0 \pmod n$. We will consider $\alpha$'s and $\beta$'s such that
$2\alpha\beta=n$ to make the sequence shorter.
The isolated equality property implies that all of the $a_i$ and $b_j$ values are distinct.
Thus, the total diversity will be $\alpha+\beta$.

\forget
\begin{lemma}
If $2\alpha\beta=0 \pmod n$, then $\alpha+\beta\geq\sqrt{2n}$.
\end{lemma}
\begin{proof}
Fix a $k$ such that $2\beta\alpha=kn$. Then $\alpha+\beta$ reduces to
$d(\alpha)=\alpha + \frac{kn}{2\alpha}$.
We wish to minimize $d(\alpha)$.
$d'(\alpha)=1-\frac{kn}{2\alpha^2}=0$ iff $\alpha=\sqrt{kn/2}$.
$d''(\alpha)=kn\alpha^{-3}>0$, thus the minimum value of $d$
is $d(\sqrt{kn/2}) =\sqrt{2kn}\ge\sqrt{2n}$.
\end{proof}
\forgotten

Under these restrictions, we can see via elementary calculus that the choice
$\alpha=\beta=\sqrt{n/2}$ yields the minimum possible
total diversity of $\alpha+\beta=\sqrt{2n}$
values.

We thus begin with $n$'s for which $n/2$ is a square, and choose $\alpha=\beta=\sqrt{n/2}$.

\forget
It is easy to verify that this meshing of disjoint $a_i$ and $b_j$
values guarantees the isolated equality property.

What is left to be shown is that for some choice of $f$ and
$x_0=s$, the sequence of actual numbers produced by the counter
assisted generator $x_i=f(x_{i-1})+i \pmod n$ has this form.
\forgotten

We now consider the specific values of the elements in our meshed sequence.
The conditions are: $c_{i,j+1}=f(c_{ij})+2\beta i+(j+1)$,
$c_{i+1,0}=f(c_{i,2\beta-1})+2\beta(i+1)-1$,
and $c_{00}=f(c_{\alpha-1,2\beta-1})+2\alpha\beta$. In terms of the $a_i$ and
$b_j$ this is:
\begin{eqnarray}
b_j & = & f(a_i)+2\beta i+(2j+1)\nonumber\\
a_i & = & f(b_j)+2\beta i+(2j+2)\quad (j=0,\ldots,\beta-2)\nonumber\\
a_i & = & f(b_{\beta-1})+2\beta i\nonumber
\end{eqnarray}
Setting $x=f(a_0)$, the first
equation yields $b_j = x+(2j+1)$ for $i=0$. Putting this back in
the equation we get that $f(a_i)=x-2\beta i$ for all $i$. \forget
a_i = f(b_j)+2\beta i+(2j+2)
-----------------------
a_i = y+2\beta i+2              /* y = f(b_0) */ => \A y-2j = f(b_j)
/* II */
\forgotten
Similarly, the second equation implies
(setting $y=f(b_0)$) $a_i = y+2\beta i+2$ and $f(b_j)=y-2j$ for all
$j<\beta-1$.
\forget
a_i = f(b_{\beta-1})+2\beta i
----------------------
y+2 = f(b_{\beta-1}) => \A i a_i = y+2\beta i+2
Putting these $a_i$'s in the third equation we get $f(b_{\beta-1})=y+2$.
\forgotten
The third equation with $i=0$ gives $f(b_{\beta-1})=a_0=y+2$.

We therefore have, for any choice of $x,y$, the following
requirements:
\begin{eqnarray}
a_i     = y+2+2\beta i & \mappedto & x-2\beta i\nonumber\\ b_j     =
x+1+2j  & \mappedto & y-2j\quad (j<\beta-1)\nonumber\\ b_{\beta-1} =
x-1+2\beta         & \mappedto & y+2\nonumber
\end{eqnarray}
It is easy to check that any such definition
yields the desired sequence of states,
as long as the resultant $a_i$ and $b_j$'s are
disjoint.
As we assume that $n$ is even, choosing any $x$ and $y$ having the
same parity (e.g., $x=y=0$) will do.

The values of $f$ on $X\setminus \{a_i,b_j\}$ can be arbitrary. It
remains to check that the sequence is repeated after every $\alpha\cdot 2\beta$ steps.
Indeed, the counter will be $2\alpha\beta=0 \pmod n$, and thus
$x_{2\alpha\beta} = f(x_{2\alpha\beta-1})+0 = f(b_{\beta-1})=a_0$, so we
are right where we begun.

\medskip

We now treat the cases where $n/2$ is not a square.
Set $\alpha = \beta =\lfloor{\sqrt{n/2}}\,\rfloor$, and define $a_i$, $b_j$, and $f$
as above. Now modify $f(x)$ to $f(x \bmod 2\alpha\beta)$. The above
argument shows that if we project the state-sequence $\x$ modulo
$2\alpha\beta$, we get diversity at most $\alpha+\beta = O(\sqrt{n})$.
Therefore, the actual diversity can be no more than $O(\sqrt{n})\cdot\lceil
n/(2\alpha\beta)\,\rceil=O(\sqrt{n})\cdot 2=O(\sqrt{n})$.
\end{proof}

\begin{remark}
In most practical cases, $n/2$ is not a square and thus we cannot achieve the exact
$\sqrt{2n}$ upper bound using our meshing construction. However, in many cases $n$
is an even power of $2$ (e.g, $2^{24}$, $2^{32}$, $2^{64}$, $2^{128}$, etc.), so
we can choose $\alpha=\sqrt{n}$ and $\beta=\sqrt{n}/2$ (note that
$2\alpha\beta=n$) to get total diversity $\alpha+\beta=3\sqrt{n}/2$, which is close
to the $\sqrt{2n}$ upper bound achieved in the case where $n/2$ was a square.
\end{remark}

\forget
\begin{lemma}
Let $n=\nu^2$  where $\nu=2^e$. Then there exists a function $f$
such that the total diversity of the counter assisted generator
$x_i=f(x_{i-1}) \xor i$ is $O(\sqrt{n})$.
\end{lemma}
\begin{proof}
Using the meshing construction, the conditions are
$c_{i,j+1}=f(c_{ij})\xor (2\nu i+(j+1))$, and
$c_{i0}=f(c_{i-1,2\nu-1})\xor (2\nu i-1)$. For all $k<2^e$,
$2k+1<2^{e+1}$. As $\nu=2^e$, we have $2\nu i+(j+1)=2\nu i \xor
(j+1)$, thus in terms of the $a_i$'s and $b_j$'s the conditions
are:
\begin{eqnarray}
b_j & = & f(a_i) \xor 2\nu i \xor (2j+1)\nonumber\\ a_i & = &
f(b_j) \xor 2\nu i \xor (2j+2)\quad (j<\nu-1)\nonumber\\ a_i & = &
f(b_{\nu-1}) \xor 2\nu i\nonumber
\end{eqnarray}
\forgotten
\forget
b_j = f(a_i) xor (2\nu i xor (2j+1))
-----------------------
b_j = x xor (2j+1)               /* x=f(a_0) */ => \A i x xor 2\nu
i = f(a_i)  /* I */

a_i = f(b_j) xor 2\nu i xor (2j+2)
-----------------------
a_i = y xor 2\nu i xor 2          /* y = f(b_0) */ => \A y xor 2j
= f(b_j)       /* II */

a_i = f(b_{\nu-1}) xor 2\nu i
----------------------
y xor 2 = f(b_{\nu-1}) => \A i a_i = y xor 2\nu i+2
\forgotten
\forget
As
before, these conditions yield for any choice of $x,y$, the
following suggestion:
\begin{eqnarray}
a_i     = y\xor 2\xor 2\nu i & \mappedto & x \xor 2\nu
i\nonumber\\ b_j     = x \xor 1\xor 2j  & \mappedto & y \xor
2j\quad (j<\nu-1)\nonumber\\ b_{\nu-1} =  x \xor 3\xor 2\nu
& \mappedto & y \xor 2\nonumber
\end{eqnarray}
which satisfies all of the above requirements, as long as the
resulted $a_i$ and $b_j$'s are disjoint (this again will be the
case if $x$ and $y$ have the same parity).

It remains to check that $\x$ enters a cycle of size $\nu\cdot
2\nu$.
$x_{\nu\cdot 2\nu}=f(x_{\nu\cdot 2\nu-1})\xor\nu\cdot 2\nu=
f(b_{\nu-1})\xor 2\nu^2=y\xor 2\xor 2n=y\xor 2=a_0=x_0$, and we
are done.
\end{proof}
\forgotten

Our construction showed that the bound $\sqrt{n}$ for the total
diversity is asymptotically tight. However, we do not have a
construction where
$\D_\G(k)$ is
$O(\sqrt{k})$ for all $k$
\emph{simultaneously}.

\begin{problem}
Does there exist a constant $c$ such that for all sufficiently large $n$,
there exists a counter-assisted generator $\G$ (with state space of size $n$)
such that $\D_\G(k)\le c\sqrt{k}$ for all $k$?
\end{problem}

\section{Cascade counter-assisted generators}
In this section we generalize the notion of counter-assisted
generators.

A Latin square is a binary function which is uniquely
invertible given its output and any one of the inputs. For
example, the operations $x+y \pmod n$, $x-y \pmod n$ and $x\xor y$ are Latin square
operations. Moreover, every group operation is a Latin square operation, and
if $x\star y$ is a Latin square operation
and $P,Q,Z$ are permutations, then $Z(P(x)\star Q(y))$ is a Latin
square operation.
Let $\star$ be a Latin square operation.%

It is easy to see that the proof of Theorem \ref{counter-assisted is good} applies when
the $+i$ modification is replaced by 
any Latin square operation $\star i$ (unique invertibility with respect to
the $i$ input guarantees the isolated equality property, and
unique invertibility with respect to the $x_i$ input guarantees the
pseudorandomness of the states). We can thus extend the concept of
counter assisted generators to include these cases as well.

\begin{remark}
When $n$ is a power of $2$, we can use essentially the same construction
as in the proof of Theorem \ref{tightness} to show the
optimality of the $\Omega(\sqrt{n})$ lower bound when the $+i \pmod n$
modification is replaced by a $\xor i$ modification.
\end{remark}

The next lemma shows that counter-mode generators are a
degenerated case of counter-assisted generators.
\begin{lemma}
Every counter-mode generator is a counter-assisted generator.
\end{lemma}
\begin{proof}
A counter-mode generator with $i$th output $g(s\star i)$ is
equivalent to the counter-assisted generator $\G=\<f,g\>$, where
$f\equiv s$, and the Latin square operation is $\star$, since in this case,
$x_i=f(x_{i-1})\star i = s\star i$.
\end{proof}


We can extend the notion of counter-assisted generators further.
Assume that $\G=\<f,g,X,Y\>$ is an iterative generator,
and let $\vec c = \<c_0,c_1,\dots\>$ be any sequence of
elements in $X$. Define the \emph{sequence-assisted generator} $\G \star \vec c$
to be the generator whose $i$th state is $x_i=f(x_{i-1})\star c_i$
(and whose $i$th output is $g(x_i)$).

\begin{theorem}\label{sequence-assisted}
Let $\G=\<f,g\>\star\vec c$ be a sequence-assisted generator. Then:
\begin{enumerate}
\Item 
{$\D_\G(k)\ge \sqrt{\D_{\vec c}(k)-1}$ for all $k=1,2,\dots$.}
\Item{If the the sequence $\vec c$ is pseudorandom,
then the state sequence of $\G$ is pseudorandom.}
\Item{If $f$ is pseudorandom, then the state sequence of $\G$ is pseudorandom.}
\end{enumerate}
\end{theorem}
\begin{proof}
(1) As in Lemma \ref{isolated-equality}, we can show that
$c_i\neq c_j$ implies $(x_{i-1},x_i)\neq (x_{j-1},x_j)$.
The rest of the proof is similar to the proof of Theorem \ref{counter-assisted is good}(1).

(2) If the state sequence of $\G$ is not pseudorandom, then the sequence $\vec c$
can be distinguished from pseudorandom noise by considering $\<f,g\>\star \vec c$,
and looking at the state sequence of $\G$.

(3) This is proved as in Theorem \ref{counter-assisted is good}(2); the only difference is
in the definition of Oracle 3.
\end{proof}

Thus, any sequence $\vec c$ with large diversity can be used instead of a counter.
In particular, we can use the output of any of the generators mentioned in Section \ref{mathgen}
as the assisting sequence. In general,
assume that $\C$ is any generator with output in $X$.
Define $\G \star \C = \G\star \vec c$, where $\vec c=\<c_0,c_1,\dots\>$ is
the output sequence of $\C$ (note that the sequence $\vec c$
depends of the initialization of $\C$).
The following definition is inductive.

\begin{definition}\label{cascade}
$\G$ is a \emph{cascade} counter-assisted generator if:
\begin{enumerate}
\item{$\G$ is a (standard) counter-assisted generator, or}
\item{$\G=\F\star\C$, where $\F$ is an iterative generator,
$\star$ is a Latin square operation, and $\C$ is a cascade
counter-assisted generator.}
\end{enumerate}
\end{definition}

In particular, we have:
\begin{lemma}
Every iterative generator is a cascade counter-assisted generator.
\end{lemma}
\begin{proof}
If $\G$ is an iterative generator, and $\C$ is a generator with
output function $0$, then $\G + \C = \G$ is a cascade
counter-assisted generator.
\end{proof}
Thus the notion of cascade counter-assisted generators extends
those of iterative, counter-mode and counter-assisted generators.

Ideally, all internal states of the cascaded generators
(including the starting position of the counter $i$)
should be initialized by random, independent seeds.
If this is not feasible, one can, e.g., initialize
the ``driving'' generator or the counter with a random seed,
and then clock the cascade a few times to
make all internal states depend on the seed. In this case,
however, caution must be taken to make sure that
particular choice of output functions does not make the influence
of the seed ``vanish'' while going
down the cascade.

\begin{example}
Assume that the generators $\A$, $\B$, and $\C$ have state
spaces of size $n=2^{256}$ ($256$ bits).
Assume further that the generator $\C$ is counter-based with
an invertible output function $g_\C$,
and that the output function $g_\B$ of $\B$ is invertible as well.
Consider the total diversity of the cascade generator $\A + \( \B\xor\C\)$ (see Figure 3):
As $\C$ is counter-based, we have $\D_\C(n)=n$. Thus by
Theorem \ref{sequence-assisted} (and discreteness),
$\D_{\B\xor\C}(n)\geq \lceil \sqrt{n-1} \rceil = 2^{128}$, and
$\D_{\A+(\B\xor\C)}(n)\geq \lceil \sqrt{\D_{\B\xor\C}(n)-1} \rceil \geq 2^{64}$.
Moreover, if the output function of $\C$, or any of the iteration functions of $\B$, $\A$ is
pseudorandom, then the state sequence of $\A$ is pseudorandom as well.
(We can also use, e.g., a maximal length LFSR instead of the
counter-based generator $\C$ to get the same
results.)
\end{example}

\begin{figure}[!h]
\begin{displaymath}
\epsfysize=5 truecm {\epsfbox{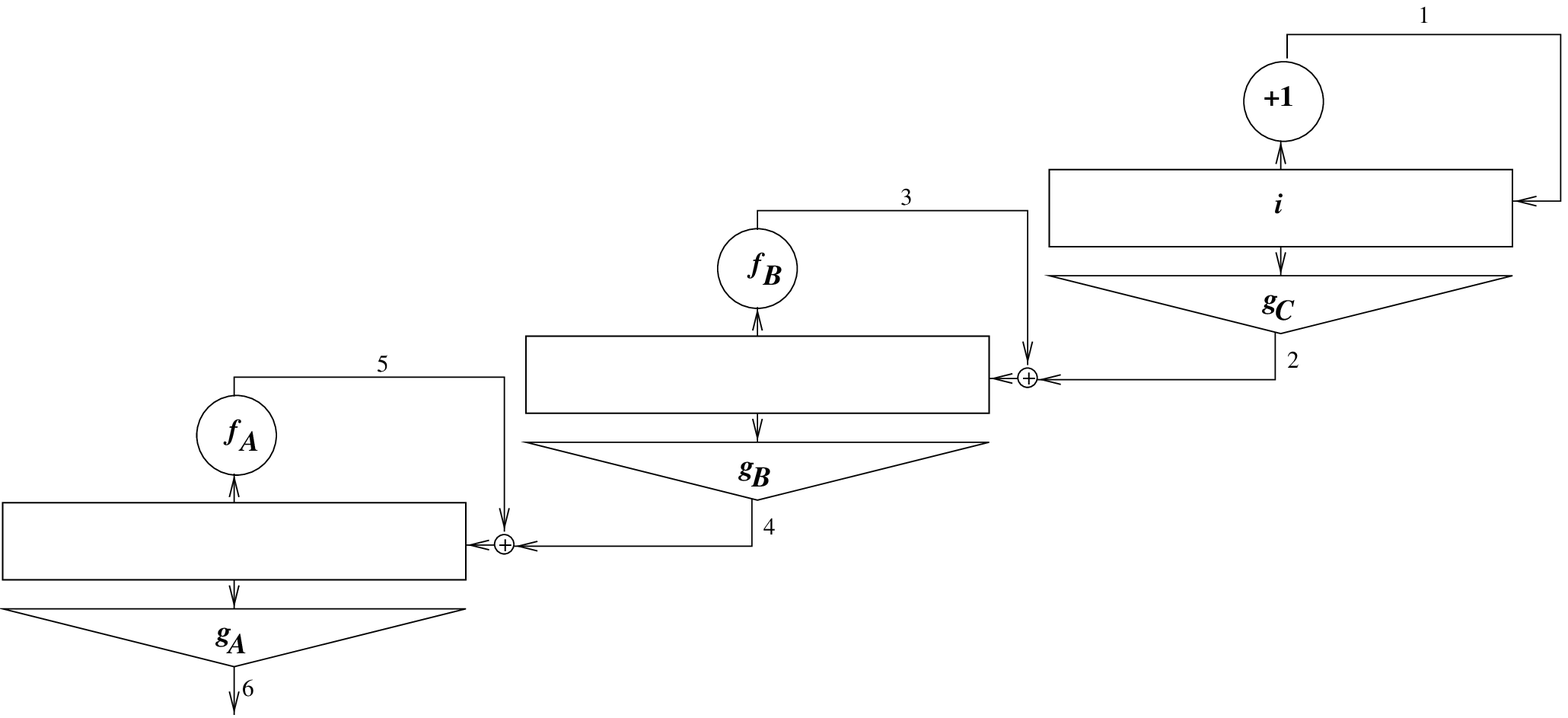}}
\end{displaymath}
\caption{}
\end{figure}

\begin{remark}
In this section we have seen that every iterative generator
can be viewed as a cascade counter-assisted
generator (in a degenerate manner). On the other hand, as
 mentioned in Section \ref{provgood}, every counter-assisted
generator can be viewed as an iterative generator
(with a larger state space). The advantage of our approach
is that we focus on the cryptographical part of the generator,
 from which the output is calculated, rather
than on the state of the whole system.
\end{remark}

\section{Generating sequences with maximal diversity}\label{maxdiv}

If we 
allow the design of a new output function $g$,
then we can modify any generator to have the maximal possible
diversity $\D_\G(k)=k$ for all $k=1,2,\ldots,n$.

\begin{definition}
Let $\G$ be any iterative generator. Modify its next-state function
as follows:
\begin{eqnarray}
x_{2i+1} &=& f(x_{2i}) \nonumber\\
x_{2i+2} &=& f(x_{2i+1})+i\nonumber
\end{eqnarray}
That is, the counter is incremented and added to the state value
only once every two iterations of the generator.
The pair of generated values $(x_{2i},x_{2i+1})$ is used as the
argument of a new output function $g':X\times X\to Y\times Y$.
We call this mode of operation \emph{the two-step counter-assisted mode}.
More generally, the \emph{$t$-step counter-assisted mode} is defined
by incrementing and adding the counter once every $t$ iterations, and using each $t$-tuple
as the input of a new output function $\hat g: X^t\to Y^t$.
Formally, the $t$-step generator $\G=\<f,g,X,Y\>$ with Latin square operation $\star i$
is the counter-assisted generator $\G^t=\<\hat f,\hat g, X^t, Y^t\>$ with the (injective)
operation $\hat \star i$, where
\begin{itemize}
\item{$\hat f(x_0,\ldots,x_{t-1}) = (f(x_{t-1}),f^2(x_{t-1}),\ldots,f^t(x_{t-1}))$,}
\item{$(x_0,\ldots,x_{t-1})\hat\star i =  (x_0,\ldots,x_{t-1}\star i)$, and}
\item{$i$ is a cyclic counter in the range $0,1,\ldots,n-1$.}
\end{itemize}
\end{definition}
Note that $t$-step counter-assisted generators require a state buffer of size $t$.

\begin{figure}[!h]
\begin{displaymath}
\epsfysize=2.9 truecm {\epsfbox{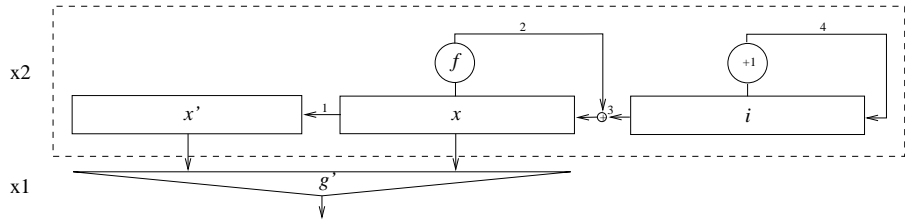}}
\end{displaymath}
\caption{A two-step counter-assisted generator}
\end{figure}


For all $t\geq 2$, any $t$-step counter-assisted generator has maximal possible diversity:

\begin{theorem}\label{fulldiver}
For any generator $\G=\<f,g\>$, and for all $t\geq 2$, we have the following:
\begin{enumerate}
\Item{If $f$ is pseudorandom, then the state sequences of $\G^t$ are pseudorandom.}
\Item{$\D_{\G^t}(k)=k$ for all $k=1,\ldots,n$.}
\end{enumerate}
\end{theorem}

\begin{proof}
The proof of the pseudorandomness part is similar to that in
Theorem \ref{counter-assisted is good}.

To prove the diversity part, assume that for some
$i\neq j\pmod n$ we have equality between the $t$-tuples $(x_{it},\ldots,x_{it+t-1})$
and $(x_{jt},\ldots,x_{jt+t-1})$. In particular,
$x_{it+t-2}=x_{jt+t-2}$. But this implies
$x_{it+t-1} = f(x_{it+t-2})+i \neq f(x_{jt+t-2})+j = x_{it+t-1} \pmod n$,
a contradiction.
\end{proof}

\subsection{Black-box modifications of the output function $g$}

If the computational complexity of evaluating the new output function
$g'$ in the two-step mode is at most double that
of evaluating $g$, then on average, the computational complexity of obtaining the next
output does not change: We clock the generator twice, but we get two outputs at once.
If the output space $Y$ is equal to $X$ then we can get very close
to this without designing a new
output function.

We will use the terminology of \cite{NAORE}. For a function $g:X\to X$, define the
\emph{Feistel permutation} $D_g : X\times X\to X\times X$ by
$D_g(L,R)\stackrel{\text{\rm def}}{=} (R, L\xor g(R))$. (Here too, any Latin square operation
$\star$ can be used instead of $\xor$.)

If the output function $g$ is key-dependent, then we can use a Luby-Rackoff construction.
Denote the key space by $K$, and assume that the size of the key space is exponential in
$\log n$.

\begin{theorem}\label{threedes}
Assume that the mapping $\k \mapsto g_\k$ is pseudorandom,
and that $\k_1$, $\k_2$, and $\k_3$ are
pseudorandom elements of $K$. Then for all functions $f: X\to X$ and seeds $x_0\in X$,
the two-step generator $\<\hat f, D_{g_{\k_1}}\circ D_{g_{\k_2}}\circ D_{g_{\k_3}}\>$
has pseudorandom output.
\end{theorem}
\begin{proof}
By Theorem \ref{fulldiver}, for all iteration functions $f$ and seeds $x_0\in X$,
the inputs to $D_{g_{\k_1}}\circ D_{g_{\k_2}}\circ D_{g_{\k_3}}$
are all distinct.
By a result of Luby and Rackoff \cite{LURA}, this implies pseudorandomness of the output.
\end{proof}

This construction makes the output calculation slower by a factor of 3:2.
The computational complexity of the following alternative is closer to the
desired optimum, and is a more straightforward modification.

\begin{theorem}\label{hashdes}
Assume that $g:X\to X$ is pseudorandom, and assume that $h:X\to X$ is pseudorandomly
chosen from a family $H$ of functions
such that for all distinct $x,y\in X$ and for all $z\in X$, the probability that
$h(x)\xor h(y)=z\ (h\in H)$ is negligible.
Then for all functions $f: X\to X$ and seeds $x_0\in X$,
the two-step counter-assisted generator $\<\hat f, D_{g}\circ D_{g}\circ D_{h}\>$
has pseudorandom output.
\end{theorem}
\begin{proof}
By a result of Lucks \cite{LUCKS} (see also \cite{NAORE}),
$D_{g}\circ D_{g}\circ D_{h}$ is pseudorandom.
The rest of the proof is like in Theorem \ref{threedes}.
\end{proof}

There exist very efficient families $H$ with the property mentioned in Theorem \ref{hashdes}
(see \cite{NAORE} for examples and references). Thus, the computational overhead of applying
$h$ is small, and the resulting generator is almost as efficient as the original one.
Note that, unlike the results in earlier sections, we get here a black-box modification of
an iterative generator $\<f,g\>$ which has maximal \emph{output} diversity, and
if \emph{either one} of the functions $f$ or $g$ is pseudorandom, then
the output sequence is pseudorandom. 

\begin{example}\label{desrc5}
Let $f=\DES$ \cite{DES}, $g=\op{RC5}$ \cite{RC5},
and $h_\k:\{0,1\}^{64}\to\{0,1\}^{64}$ be a function
from Vazirani's \emph{shift family} (the $i$th bit
of $h_\k(x)$ is $\sum_{i=1}^n x_i \k_{j+i-1} \bmod 2$,
see \cite{NAORE} and \cite{VAZIRANI}).
The two-step counter-assisted generator
$\<\widehat{\DES}, D_{\op{RC5}}\circ D_{\op{RC5}}\circ D_{h_\k}\>$
has maximal (state and output) diversity $k$ for all $k=1,2,\ldots,2^{64}$.
On average, the calculation of any output $64$
 bit block requires a single invocation of $\DES$ and a single invocation
of $\op{RC5}$. The execution time overhead of the rest of the operations is negligible.
Furthermore, if \emph{either one of the two} functions
$\DES$ and $\op{RC5}$ is difficult to distinguish from
random, then the output sequence will be difficult to distinguish from random as well.
\end{example}

\begin{problem}
Assume that both $f$ and $g$ are (truly) random,
and consider an output sequence of length $m$ generated from a random seed by the
two-step counter-assisted generator $\G^2=\<\hat f,D_g\circ D_g\>$.
What is the highest distinguishing probability between such a sequence and a random sequence?
\end{problem}

\begin{remark}
Using the results from \cite{NAORE}, we get that for
all $t$, the output function of the $t$-step counter-assisted mode
can be modified in a black-box manner with a small computational overhead, to get the same
diversity and pseudorandomness results. See \cite{NAORE} for details.
\end{remark}

\begin{remark}
In certain cases, when $t$ is large (e.g., $t\geq 4$) it is desirable that the inputs to the
$t$-step output function are distinct in as many entries
as possible (for example, this guarantees many active $S$-boxes in differential cryptanalysis of
the output function). We can achieve this goal via letting the next state be the same as when
clocking the (standard) counter-assisted generator $t$ times (that is, the counter is incremented
and added to the $x_i$ value every clock).
By the isolated equality property, this guarantees that
any two $t$-tuples are distinct in at least $\lfloor t/2\rfloor$ entries.
In this mode of operation, the diversity remains maximal as long as $k< n/t$.
\end{remark}

\subsection{Safe transition to new generations of cryptographic functions}
A common practice in the design of new generations of cryptographic functions is to double the
input and output length. Nowadays, we experience the evolution from $64$ bit
functions (such as DES, RC5, etc.) to $128$ bit functions (such as the
AES candidates \cite{AES}).
The advantage of old generation functions is that they have gone through
years of extensive academic research, and are thus well understood. It will
take a long time to gain similar confidence in the new generation functions.

Our two-step counter-assisted mode suggests a natural and straightforward
way to combine new and old generation functions in a way that if \emph{either one} of
them is pseudorandom, then the resulting generator is pseudorandom:
Assume that $f$ is an old generation function and $g$ is a new generation function
with double input size. Then we simply use the two-step counter-assisted
generator $\<\hat f, g\>$.

\begin{example}\label{AES}
In Example \ref{desrc5}, we can use RC6 instead of
$D_{\op{RC5}}\circ D_{\op{RC5}}\circ D_{h_\k}$
as the output function. This results in a faster and more elegant generator.
Here too, the diversity is maximal for all $k=1,\ldots,2^{64}$, and the generator is
difficult to distinguish from random if either DES or RC6 is.
\end{example}

\subsection{Cascaded multiple-step counter-assisted generators}

If we have enough state-space (this is usually the case with software encryption),
we can cascade multiple-step counter-assisted generators without decreasing the
diversity.
Consider for example generators $\G_0,\G_1,\ldots,\G_{m-1}$
having the same state-space and output-space.
For any sequence of positive integers $t_0 < t_1 <  \ldots <  t_{m-1}$, and
Latin-square operations $\star_{t_0},\ldots,\star_{t_{m-2}}$ (on spaces of
size $t_0, t_1, \ldots, t_{m-2}$ blocks, respectively),
the \emph{$(t_0, t_1, \ldots, t_{m-1})$-step cascade} is defined to be
$$\G_{\textrm{\rm cascade}}=\G_{m-1}^{t_{m-1}}\hat\star_{t_{m-2}}%
\ldots\hat\star_{t_1}\G_1^{t_1}\hat\star_{t_0}\G_0^{t_0}.$$
In the sense of definition \ref{cascade}.
Here,
$(x_0,\ldots,x_{t_{j+1}-1})\hat\star_{t_j} (y_0,\ldots,y_{t_j-1})$
is defined as the concatenation of
$(x_0,\ldots,x_{t_{j+1}-t_j-1})$ and
$(x_{t_{j+1}-t_j},\ldots,x_{t_{j+1}-1})\star_{t_j}(y_0,\ldots,y_{t_j-1})$.

Using this notation, we have the following:
\begin{theorem} For all generators $\G_0,\G_1,\ldots,\G_{m-1}$
having the same state-space and output-space,
and for any Latin-square operations $\star_{t_0},\ldots,\star_{t_{m-2}}$ (on spaces of
size $t_0 < t_1 < \ldots < t_{m-2}$ blocks, respectively),
the \emph{$(t_0, t_1, \ldots, t_{m-1})$-step cascade}
$\G_{\textrm{\rm cascade}}=\G_{m-1}^{t_{m-1}}\hat%
\star_{t_{m-2}}\ldots\hat\star_{t_1}\G_1^{t_1}\hat%
\star_{t_0}\G_0^{t_0}$ has the following properties:
\begin{enumerate}
\Item{$\D_{\G_{\textrm{\rm cascade}}}(k)=k$ for all $k=1,2,\ldots n$.}
\Item{If either the iteration or the output function of any of the cascaded generators is
pseudorandom, then the output of $\G_{\textrm{\rm cascade}}$ is pseudorandom as well.}
\end{enumerate}
\end{theorem}
\begin{proof}
(1) follows from Theorem \ref{fulldiver}, by induction on $m$.
(2) follows readily from Theorem \ref{sequence-assisted}.
\end{proof}


\section{Concluding remarks and further research}

We have presented a new mode of operation which makes the diversity of every state sequence
provably large with a negligible computational cost.
Unlike other solutions, this mode does not introduce new (trivial) risks.
The well known threat of ``no available theory'' on the cycle structure of
complicated iterative generators
(see, e.g., \cite[p.~525]{CHAMGOL}, \cite[p.~22]{CHAMBERS}, \cite[\S 17.6]{S}, and \cite[p.~347]{CDR}) is eliminated.
It is important to stress, however, that the diversity measures only one aspect of
security, and is clearly not sufficient for evaluating the
cryptographical strength of the generator.

Our new mode has various possible implementations via multiple-stepping and/or cascading, which
allow the user a wide range of choice to fit the implementation to his constraints and needs.
All of the suggested modes require a counter, but in most of the applications a counter either
already exists or is easy to maintain. The cascaded mode reduces the provable diversity
with respect to the simple counter-assisted mode, but it suggests an interesting new way
to combine the cryptographic strength of several generators. The multiple-stepping mode requires
a larger state buffer (thus may be more suitable in software applications), but
assures perfect diversity.

The cryptographical impact of our modification technique
when the functions $f$ or $g$ are not pseudorandom remains
open. It is easy to find pathological examples
of output functions where the modification makes things worse,
but we believe that such pathological cases will be easy to inspect.
However, if the user wants complete confidence, then he may wish to replace the output function
$g$ by one that he trusts. In this case, it may be worthwhile to use the generator in the
two-step mode and gain the maximal possible diversity as in Section \ref{maxdiv}.

As we have proved, in the multiple-stepping modes it is enough
that either the iteration \emph{or} the output function is pseudorandom to
obtain pseudorandom output. This suggests combining two functions from ``orthogonal''
sources, such as in Example \ref{desrc5}, and combining strength of well studied primitives
with with new, promising ones, as in Example \ref{AES}.

The counter-assisted mode suggests many open problems. Some of these problems are mentioned
in the paper. To these we can add practical problems such as the challenge of finding a seed $s$
for which the counter-assisted generator with DES as the iteration function has
$\D_{DES(s)}(k)\thickapprox \sqrt{k}$ for some large $k$, and theoretical
problems such as statistical analysis of the behavior of the
state sequence of counter-assisted generators.

\end{document}